\documentclass[seceq]{ptptex}
\usepackage{wrapft}
\usepackage{bm}

\title{
Equivalence Classes of Boundary Conditions in $SU(N)$ Gauge Theory on 2-dimensional Orbifolds
}

\author{
Yoshiharu \textsc{Kawamura},\footnote{E-mail: haru@azusa.shinshu-u.ac.jp} 
and Takashi \textsc{Miura}
}

\inst{
Department of Physics, Shinshu University, Matsumoto 390-8621, Japan
}

\recdate{
May 26, 2009}

\abst{
We study equivalence classes of boundary conditions 
in an $SU(N)$ gauge theory on six-dimensional space-time including two-dimensional orbifold.
For five kinds of two-dimensional orbifolds $S^1/Z_2 \times S^1/Z_2$ and $T^2/Z_m$ $(m=2,3,4,6)$, 
orbifold conditions and those gauge transformation properties are given 
and the equivalence relations among boundary conditions are derived.
The classification of boundary conditions related to diagonal representatives is carried out 
using the equivalence relations.
}

\begin{document}

\maketitle

\section{Introduction}

Grand unified theories on an orbifold have been attracted phenomenologically
since Higgs mass splitting was well realized by the orbifold breaking mechanism.\cite{Kawamura1,Hall1}$^,$\footnote{
In four-dimensional heterotic string models, extra colored Higgs are projected by the Wilson line mechanism.\cite{Higgs}}
Various kinds of models have been constructed and
those come from the variety of choice for ingredients such as gauge groups, representations of fields,
extra dimensions and boundary conditions (BCs) for fields.
The features of the first three ingredients have been studied intensively, 
but those of the last one have not been fully understood with a few exceptions such as 
BCs on the orbifolds $S^1/Z_2$, $T^2/Z_2$ and $T^2/Z_3$.

The BCs for bulk fields are classified into 
the equivalence classes using the gauge invariance.
Several sets of BCs belong to the same equivalence class and describe the same physics,
if they are related to gauge transformations.
Specifically, the symmetry of BCs is not necessarily the same as the physical symmetry.
The physical symmetry is determined by the Hosotani mechanism after the rearrangement of gauge symmetry.\cite{H}
Equivalence classes of BCs and dynamical gauge symmetry breaking were studied for gauge theories
on $S^1/Z_2$\cite{HHHK,HHK}$^,$\footnote{
See Ref. \citen{KLY} for the breakdown of gauge symmetry on $S^1/Z_2$ by the Hosotani mechanism.}
$T^2/Z_2$\cite{HN&T} and $T^2/Z_3$\cite{KK&M}.
It is interesting to study equivalence classes of BCs for gauge theories on other orbifolds 
and to construct a phenomenologically viable model based on them.

In the present paper, we study equivalence classes of BCs 
in an $SU(N)$ gauge theory on six-dimensional space-time including two-dimensional orbifold.
For five kinds of two-dimensional orbifolds $S^1/Z_2 \times S^1/Z_2$ and $T^2/Z_m$ $(m=2,3,4,6)$, 
orbifold conditions and those gauge transformation properties are given 
and the equivalence relations among BCs are derived.
The classification of BCs related to diagonal representatives is carried out using the equivalence relations.

In \S 2, general arguments are given for BCs in gauge theories on $S^1/Z_2 \times S^1/Z_2$.
Equivalence classes of BCs are defined by the gauge invariance
and the classification of BCs is carried out using the equivalence relations among BCs. 
In \S 3, we study the equivalence classes of BCs
and classify BCs related to diagonal representatives on $T^2/Z_m$ $(m=2,3,4,6)$.
Section 4 is devoted to conclusions.

\section{$S^1/Z_2\times S^1/Z_2$ Orbifold and Equivalence classes}

\subsection{Boundary conditions}

We study an $SU(N)$ gauge theory defined on a six-dimensional space-time 
$M^4\times S^1/Z_2\times S^1/Z_2$\footnote{
The models on $M^4\times S^1/Z_2\times S^1/Z_2$ were studied in Ref.~\citen{Li}.
} 
where $M^4$ is the four-dimensional Minkowski space-time. 
An extra space $S^1/Z_2\times S^1/Z_2$ is obtained by identifying points on $T^2 = S^1 \times S^1$ by parity.
Let $x$ and $\vec{y}=(y_1,y_2)$ be coordinates of $M^4$ and $S^1/Z_2\times S^1/Z_2$, respectively.
On $S^1\times S^1$, the points $\vec{y}+\vec{e}_1$ and $\vec{y}+\vec{e}_2$
are identified with the point $\vec{y}$
where $\vec{e}_1$ and $\vec{e}_2$ are basis vectors and
we take them as the unit vectors $\vec{e}_1 = (1,0)$ and $\vec{e}_2 = (0, 1)$.\footnote{
On the estimation of physical quantities, we use physical sizes 
such as $|\vec{e}_1| = 2\pi R_1$ and $|\vec{e}_2| = 2 \pi R_2$.
We use the integral multiple of unit vector as basis vectors for other orbifolds.}
The orbifold $S^1/Z_2\times S^1/Z_2$ is obtained by further identifying 
$(-y_1,y_2)$ and $(y_1,-y_2)$ by $(y_1,y_2)$. 
The fixed lines or points on $S^1/Z_2\times S^1/Z_2$ are lines or points that transform themselves
under the $Z_2$ transformations $\vec{y} \rightarrow \theta_1\vec{y} = (-y_1, y_2)$, 
$\vec{y} \rightarrow \theta_2\vec{y} = (y_1, -y_2)$ or 
$\vec{y} \rightarrow \theta_3\vec{y} (=\theta_1\theta_2\vec{y} =\theta_2\theta_1\vec{y})= (-y_1, -y_2)$.
There are two fixed lines $(0, y_2)$ and $(1/2, y_2)$ for the first $Z_2$ transformation.
There are two fixed lines $(y_1, 0)$ and $(y_1, 1/2)$ for the second $Z_2$ transformation.
There are four fixed points $\vec{0}(=(0,0))$, ${\vec{e}_1/ 2}$, ${\vec{e}_2/ 2}$ 
and ${(\vec{e}_1+\vec{e}_2)/2}$ for the third $Z_2$ transformation.
Around these lines and points, we define following ten kinds of transformations:
\begin{align}
s_{10}&:\vec{y}\rightarrow \theta_1\vec{y},~~
s_{11}:\vec{y}\rightarrow \theta_1\vec{y}+\vec{e}_1,~~ 
s_{20}:\vec{y}\rightarrow \theta_2\vec{y},~~
s_{21}:\vec{y}\rightarrow \theta_2\vec{y}+\vec{e}_2,\notag \\
s_{30}&:\vec{y}\rightarrow \theta_3\vec{y},~~
s_{31}:\vec{y}\rightarrow \theta_3\vec{y}+\vec{e}_1,~~ 
s_{32}:\vec{y}\rightarrow \theta_3\vec{y}+\vec{e}_2, \notag \\
s_{33}&:\vec{y}\rightarrow \theta_3\vec{y}+\vec{e}_1+\vec{e}_2,~~
t_1:\vec{y}\rightarrow \vec{y}+\vec{e}_1,~~
t_2:\vec{y}\rightarrow \vec{y}+\vec{e}_2.
\end{align}
These satisfy the following relations:
\begin{align}
&s_{10}^2=s_{11}^2=s_{20}^2=s_{21}^2=s_{30}^2=s_{31}^2=s_{32}^2=s_{33}^2=I ,~~ \notag \\
&s_{11}=t_1s_{10},~~ s_{21}=t_2s_{20},~~ t_1t_2=t_2t_1, \notag \\
&s_{30}=s_{10}s_{20}=s_{20}s_{10},~~ s_{31}=s_{11}s_{20}=s_{20}s_{11}, \notag \\
& s_{32}=s_{10}s_{21}=s_{21}s_{10}, ~~ s_{33}=s_{11}s_{21}=s_{21}s_{11},
\label{relS1/Z22}
\end{align}
where $I$ is the identity operation.
On $S^1/Z_2\times S^1/Z_2$, the points $\vec{y}$ is identified by $\vec{y}+\vec{e}_i$ $(i=1,2)$ 
and $\theta_j\vec{y}$ $(j=1,2,3)$,
but all six-dimensional bulk fields do not necessarily take identical vaules at these points.
Under the requirement that the Lagrangian density should be single-valued on $M^4 \times S^1/Z_2\times S^1/Z_2$, 
the following BCs for gauge field $A_M(x,\vec{y})$ are allowed,
\begin{align}
s_{10}:~&A_M(x,\theta_1\vec{y})=\kappa_{[M]}^{10} P_{10}A_M(x,\vec{y})P_{10}^{\dag} ,\notag \\ 
&\hspace{4cm}\mathrm{for}~\kappa_{[\mu]}^{10}=1,~\kappa_{[y_1]}^{10}=-1,~\kappa_{[y_2]}^{10}=1, 
\label{AM-s10} \\
s_{11}:~&A_{M}(x,\theta_1\vec{y}+\vec{e}_1)=\kappa_{[M]}^{11}P_{11}A_{M}(x,\vec{y})P_{11}^{\dag}, \notag \\
&\hspace{4cm}\mathrm{for}~\kappa_{[\mu]}^{11}=1,~\kappa_{[y_1]}^{11}=-1,~\kappa_{[y_2]}^{11}=1, \\
s_{20}:~&A_M(x,\theta_2\vec{y})=\kappa_{[M]}^{20} P_{20}A_M(x,\vec{y})P_{20}^{\dag}, \notag \\ 
&\hspace{4cm}\mathrm{for}~\kappa_{[\mu]}^{20}=1,~\kappa_{[y_1]}^{20}=1,~\kappa_{[y_2]}^{20}=-1, \\
s_{21}:~&A_{M}(x,\theta_2\vec{y}+\vec{e}_2)=\kappa_{[M]}^{21}P_{21}A_{M}(x,\vec{y})P_{21}^{\dag}, \notag \\
&\hspace{4cm}\mathrm{for}~\kappa_{[\mu]}^{21}=1,~\kappa_{[y_1]}^{21}=1,~\kappa_{[y_2]}^{21}=-1, 
\label{AM-s21} \\
s_{30}:~&A_{M}(x,\theta_3\vec{y})=\kappa_{[M]}^{30}P_{30}A_{M}(x,\vec{y})P_{30}^{\dag}, \notag \\
&\hspace{4cm}\mathrm{for}~\kappa_{[\mu]}^{30}=1,~\kappa_{[y_1]}^{30}=-1,~\kappa_{[y_2]}^{30}=-1, \\
s_{31}:~&A_{M}(x,\theta_3\vec{y}+\vec{e}_1)=\kappa_{[M]}^{31}P_{31}A_{M}(x,\vec{y})P_{31}^{\dag}, \notag \\
&\hspace{4cm}\mathrm{for}~\kappa_{[\mu]}^{31}=1,~\kappa_{[y_1]}^{31}=-1,~\kappa_{[y_2]}^{31}=-1, \\
s_{32}:~&A_{M}(x,\theta_3\vec{y}+\vec{e}_2)=\kappa_{[M]}^{32}P_{32}A_{M}(x,\vec{y})P_{32}^{\dag}, \notag \\
&\hspace{4cm}\mathrm{for}~\kappa_{[\mu]}^{32}=1,~\kappa_{[y_1]}^{32}=-1,~\kappa_{[y_2]}^{32}=-1, \\
s_{33}:~&A_{M}(x,\theta_3\vec{y}+\vec{e}_1+\vec{e}_2)=\kappa_{[M]}^{33}P_{33}A_{M}(x,\vec{y})P_{33}^{\dag}, \notag \\
&\hspace{4cm}\mathrm{for}~\kappa_{[\mu]}^{33}=1,~\kappa_{[y_1]}^{33}=-1,~\kappa_{[y_2]}^{33}=-1, \\
t_1:~&A_{M}(x,\vec{y}+\vec{e}_1)=U_1A_{M}(x,\vec{y})U_1^{\dag},\\
t_2:~&A_{M}(x,\vec{y}+\vec{e}_2)=U_2A_{M}(x,\vec{y})U_2^{\dag},
\label{AM-t2}
\end{align}
where $P_{10},~P_{11},~P_{20},~P_{21},~P_{30},~P_{31},~P_{32},~P_{33},~U_1$ and $U_2$ are $N\times N$ matrices.
Here, we take matrices with constant elements to define the BCs for the bulk fields, for simplicity.
The counterparts of Eq.(\ref{relS1/Z22}) are given by
\begin{align}
&P_{10}^2=P_{11}^2=P_{20}^2=P_{21}^2=P_{30}^2=P_{31}^2=P_{32}^2=P_{33}^2=I,\notag \\
&P_{11}=U_1P_{10},~~ P_{21}=U_2P_{20},~~ U_1U_2=U_2U_1,\notag \\
&P_{30}=P_{10}P_{20}=P_{20}P_{10},~~ P_{31}=P_{11}P_{20}=P_{20}P_{11},\notag \\
&P_{32}=P_{10}P_{21}=P_{21}P_{10},~~ P_{33}=P_{11}P_{21}=P_{21}P_{11} ,
\end{align}
where $I$ stands for the $N\times N$ unit matrix.
Then the BCs in $SU(N)$ gauge theories on $S^1/Z_2\times S^1/Z_2$ are 
specified with $(P_{10},P_{11}$, $P_{20},P_{21},P_{30},P_{31},P_{32},P_{33},U_1,U_2)$.
Because four of them are independent,
we choose four kinds of unitary and hermitian matrices $P_{10}$, $P_{11}$, $P_{20}$ and $P_{21}$ as independent ones
and often refer to them simply as {\it BCs}.

\subsection{Gauge invariance and equivalence class}

Given the BCs $(P_{10},P_{11},P_{20},P_{21})$, there still remains residual gauge invariance.
Under gauge transformation with the transformation function $\Omega(x,\vec{y})$,
$A_M$ is transformed as
\begin{align}
A_M \to A'_M = \Omega A_M \Omega^{\dagger} - \frac{i}{g}\Omega \partial_M \Omega^{\dagger} ,
\end{align}
where $A'_M$ satisfies, instead of Eqs. (\ref{AM-s10}) -- (\ref{AM-s21}),
\begin{align}
s_{10}:~&A'_M(x,\theta_1\vec{y})=\kappa_{[M]}^{10} \left(P'_{10}A'_M(x,\vec{y}){P'}_{10}^{\dag} 
 - \frac{i}{g} P'_{10} \partial_M {P'}_{10}^{\dagger}\right) , \\
s_{11}:~&A'_{M}(x,\theta_1\vec{y}+\vec{e}_1)=\kappa_{[M]}^{11} \left(P'_{11}A'_{M}(x,\vec{y}){P'}_{11}^{\dag}
 - \frac{i}{g} P'_{11} \partial_M {P'}_{11}^{\dagger}\right) , \\
s_{20}:~&A'_M(x,\theta_2\vec{y})=\kappa_{[M]}^{20} \left(P'_{20}A'_M(x,\vec{y}){P'}_{20}^{\dag}
 - \frac{i}{g} P'_{20} \partial_M {P'}_{20}^{\dagger}\right) , \\
s_{21}:~&A'_{M}(x,\theta_2\vec{y}+\vec{e}_2)=\kappa_{[M]}^{21} \left(P'_{21}A'_{M}(x,\vec{y}){P'}_{21}^{\dag}
  - \frac{i}{g} P'_{21} \partial_M {P'}_{21}^{\dagger}\right) .
\end{align}
Here  $P'_{10}$, $P'_{11}$, $P'_{20}$ and $P'_{21}$ are given by 
\begin{align}
&P'_{10}(\vec{y})=\Omega(x,\theta_1\vec{y})P_{10}\Omega^{\dag}(x,\vec{y}) , ~~
P'_{11}(\vec{y})=\Omega(x,\theta_1\vec{y}+\vec{e}_1)P_{11}\Omega^{\dag}(x,\vec{y}) ,\notag \\
&P'_{20}(\vec{y})=\Omega(x,\theta_2\vec{y})P_{20}\Omega^{\dag}(x,\vec{y}) , ~~
P'_{21}(\vec{y})=\Omega(x,\theta_2\vec{y}+\vec{e}_2)P_{21}\Omega^{\dag}(x,\vec{y}) .
\end{align} 

Theories with different BCs should be equivalent in terms of physics content if they are connected by 
gauge transformations.
The key observation is that physics should not depend on the gauge chosen.
The equivalence is guaranteed in the Hosotani mechanism\cite{H}
and the two sets of BCs are equivalent:
\begin{align}
(P_{10},P_{11},P_{20},P_{21}) \sim (P'_{10}(\vec{y}),P'_{11}(\vec{y}),P'_{20}(\vec{y}),P'_{21}(\vec{y})) .
\label{Eq-rel}
\end{align}
The corresponding relations for $P'_{10}$, $P'_{11}$, $P'_{20}$ and $P'_{21}$ are given by
\begin{align}
& {P'}_{10}(\vec{y}){P'}_{10}(\theta_1\vec{y})={P'}_{10}(\theta_1\vec{y}){P'}_{10}(\vec{y})=I , \notag \\
& {P'}_{11}(\vec{y}){P'}_{11}(\theta_1\vec{y}+\vec{e}_1)={P'}_{11}(\theta_1\vec{y}+\vec{e}_1){P'}_{11}(\vec{y})=I , \notag \\
& {P'}_{20}(\vec{y}){P'}_{20}(\theta_2\vec{y})={P'}_{20}(\theta_2\vec{y}){P'}_{20}(\vec{y})=I , \notag \\
& {P'}_{21}(\vec{y}){P'}_{21}(\theta_2\vec{y}+\vec{e}_2)={P'}_{21}(\theta_2\vec{y}+\vec{e}_2){P'}_{21}(\vec{y})=I .
\end{align} 
In the case that $P'_{10}$, $P'_{11}$, $P'_{20}$ and $P'_{21}$ are independent of $\vec{y}$, 
the above relations reduce to the usual ones ${P'}_{10}^2={P'}_{11}^2={P'}_{20}^2={P'}_{21}^2=I$.
The equivalence relation (\ref{Eq-rel}) defines equivalence classes of the BCs.

We illustrate the change of BCs under a singular gauge transformation, by using
an $SU(2)$ gauge theory with the gauge transformation function defined by
\begin{align}
\Omega(\vec{y})=\exp[i\alpha(a\tau_1+b\tau_2)y_1+i\beta(a\tau_1+b\tau_2)y_2]   ~~~ (\alpha, \beta, a, b \in \mathbb{R}),
\label{Omega1}
\end{align}
where $\tau_k~(k=1,2,3)$ are Pauli matrices.
When we take $(P_{10}, P_{11})= (\tau_3, \tau_3)$, they are transformed as  
\begin{align}
& P'_{10}=\Omega(x,\theta_1\vec{y})P_{10}\Omega^{\dag}(x,\vec{y}) = \exp[i\beta(a\tau_1+b\tau_2)y_2]\tau_3 , \\
& P'_{11}=\Omega(x,\theta_1\vec{y}+\vec{e}_1)P_{11}\Omega^{\dag}(x,\vec{y}) = \exp[i\alpha(a\tau_1+b\tau_2) + 
i\beta(a\tau_1+b\tau_2)y_2]\tau_3 .
\end{align} 
$P'_{10}$ becomes diagonal with $\beta = 0$ and then $P'_{10}$ and $P'_{11}$ take the following form:
\begin{align}
&P'_{10}=\tau_3 , \\
&P'_{11}=\exp[i(a\tau_1+b\tau_2)]\tau_3 \notag \\
&~~~~= \left(I\cos\sqrt{a^2 + b^2} + i\frac{a\tau_1 + b\tau_2}{\sqrt{a^2+b^2}}\sin\sqrt{a^2+b^2}\right) \tau_3 ,
\end{align} 
where we set $\alpha = 1$ and $I$ is the $2 \times 2$ unit matrix.
$P'_{11}$ also becomes the diagonal form $(-1)^n \tau_3$ when $\sqrt{a^2 + b^2} = n\pi$ with an integer $n$.
In order to obtain a diagonal representative for both $(P_{20},P_{21})$ and $(P'_{20},P'_{21})$
with the gauge transformation function $\Omega(\vec{y})=\exp[i(a\tau_1+b\tau_2)y_1]$, 
$P_{20}$ and $P_{21}$ should be $I$ or $-I$.
In this way, we find the following equivalence relation:
\begin{align}
(\tau_3, \tau_3, \eta_{20} I, \eta_{21} I) 
\sim (\tau_3, e^{i(a\tau_1+b\tau_2)}\tau_3, \eta_{20} I, \eta_{21} I) ,
\label{R1}
\end{align}
where $\eta_{20}$ and $\eta_{21}$ take $1$ or $-1$.
In the same way, we obtain the following equivalence relation:
\begin{align}
(\eta_{10} I, \eta_{11} I, \tau_3, \tau_3) 
\sim  (\eta_{10} I, \eta_{11} I, \tau_3, e^{i(a\tau_1+b\tau_2)}\tau_3) ,
\label{R2}
\end{align}
where $\eta_{10}$ and $\eta_{11}$ take $1$ or $-1$.
The equivalence relations between diagonal representatives are given by
\begin{align}
& (\tau_3, \tau_3, \eta_{20} I, \eta_{21} I) 
\sim (\tau_3, -\tau_3, \eta_{20} I, \eta_{21} I) ,
\label{R1-d}\\
& (\eta_{10} I, \eta_{11} I, \tau_3, \tau_3) 
\sim  (\eta_{10} I, \eta_{11} I, \tau_3, -\tau_3) .
\label{R2-d}
\end{align}

\subsection{Classification of boundary conditions}

We carry out the classification of BCs for bulk fields on the orbifold 
$S^1/Z_2 \times S^1/Z_2$.

First we show that all BCs are specified by diagonal matrices for $SU(2)$ gauge group.
(1) In the case that $P_{10}$, $P_{11}$ and $P_{20}$ are the $2 \times 2$ unit matrix $I$ up to a sign factor,
$P_{21}$ can be diagonalized by global $SU(2)$ transformation.
(2) In the case that $P_{10}$ and  $P_{11}$ are $I$ up to a sign factor
and $P_{20}$ is a non-diagonal form, we derive $P_{20} = \pm \tau_3$ after global $SU(2)$ transformation.
Then $P_{21} = \pm \tau_3 \exp[i(a\tau_1 + b\tau_2)]$ is allowed, 
but we obtain $P_{21} = \pm \tau_3$ by the gauge transformation with $\Omega(x, \vec{y}) = \exp[i(a\tau_1 + b\tau_2)y_2]$.
(3) In the case that $P_{10}$ is $\pm I$ and $P_{11}$ is a non-diagonal form, 
we derive $P_{11} = \pm \tau_3$ after global $SU(2)$ transformation.
We obtain $P_{20} = \pm I$ or $\pm \tau_3$ and $P_{21} = \pm I$ or $\pm \tau_3$
using the relation $P_{11}P_{20} = P_{20}P_{11}$ and $P_{11}P_{21} = P_{21}P_{11}$, respectively.
(4) In the case that $P_{10}$ is a non-diagonal form, we derive $P_{10} = \pm \tau_3$ after global $SU(2)$ transformation.
We obtain $P_{20} = \pm I$ or $\pm \tau_3$ and $P_{21} = \pm I$ or $\pm \tau_3$
using the relation $P_{10}P_{20} = P_{20}P_{10}$ and $P_{10}P_{21} = P_{21}P_{10}$, respectively.
If $P_{20}$ or $P_{21}$ is $\pm \tau_{3}$, $P_{11} = \pm I$ or $\pm \tau_3$ using the relation 
$P_{11}P_{20} = P_{20}P_{11}$ or $P_{11}P_{21} = P_{21}P_{11}$.
If both $P_{20}$ and $P_{21}$ are $\pm I$, $P_{11} = \pm \tau_3 \exp[i(a\tau_1 + b\tau_2)]$ is allowed.
In this case, we obtain $P_{11} = \pm \tau_3$ after the gauge transformation with
$\Omega(x, \vec{y}) = \exp[i(a\tau_1 + b\tau_2)y_1]$.

In the similar way, all BCs for $SU(N)$ gauge group are made ones specified by diagonal matrices
after suitable global unitary transformations and local gauge transformations on $S^1/Z_2 \times S^1/Z_2$.
We sketch the proof.
$P_{10}$ and $P_{11}$ can be diagonalized by a global unitary transformation and a local gauge transformation
using the same argument as the case with $S^1/Z_2$.\cite{HHK}
Note that $P_{10}$ and $P_{11}$ remain $\vec{y}$-independent after the transformations 
as shown from the equivalent relation (\ref{R1}).
{}From the relations $P_{10}P_{20} = P_{20}P_{10}$, $P_{10}P_{21} = P_{21}P_{10}$, $P_{11}P_{20} = P_{20}P_{11}$
and $P_{11}P_{21} = P_{21}P_{11}$, we find that $P_{20}$ and $P_{21}$ are block diagonal matrices
and $P_{20}$ is diagonalized by a unitary matrix which belongs to a subgroup of $SU(N)$.
Then $P_{21}$ can be also diagonalized by a suitable gauge transformation following the equivalent relation (\ref{R2}).
In this way, we find that there are at least one diagonal representatives of BCs in every equivalent class.
The diagonal $P_{10}$, $P_{11}$, $P_{20}$ and $P_{21}$ in $SU(N)$ gauge theories are specified 
by sixteen non-negative integers $(p_k,q_k,r_k,s_k)$ $(k=1,2,3,4)$ such that
\begin{align}
&P_{10} = {\mbox{diag}}(\overbrace{[+1]_{p_1},[+1]_{p_2},[+1]_{p_3},[+1]_{p_4}}^p,
\overbrace{[+1]_{q_1},[+1]_{q_2},[+1]_{q_3},[+1]_{q_4}}^q,\notag \\
&\hspace{2.5cm}\overbrace {[-1]_{r_1},[-1]_{r_2},[-1]_{r_3},[-1]_{r_4}}^r,
\overbrace{[-1]_{s_1},[-1]_{s_2},[-1]_{s_3},[-1]_{s_4}}^{N-p-q-r}) ,\nonumber \\
&P_{11} = {\mbox{diag}}([+1]_{p_1},[+1]_{p_2},[+1]_{p_3},[+1]_{p_4},[-1]_{q_1},[-1]_{q_2},[-1]_{q_3},[-1]_{q_4},\notag \\
&\hspace{2.5cm}[+1]_{r_1},[+1]_{r_2},[+1]_{r_3},[+1]_{r_4},[-1]_{s_1},[-1]_{s_2},[-1]_{s_3},[-1]_{s_4}) ,\nonumber \\
&P_{20} = {\mbox{diag}}([+1]_{p_1},[+1]_{p_2},[-1]_{p_3},[-1]_{p_4},[+1]_{q_1},[+1]_{q_2},[-1]_{q_3},[-1]_{q_4},\notag \\
&\hspace{2.5cm}[+1]_{r_1},[+1]_{r_2},[-1]_{r_3},[-1]_{r_4},[+1]_{s_1},[+1]_{s_2},[-1]_{s_3},[-1]_{s_4}) ,\nonumber \\
&P_{21} = {\mbox{diag}}([+1]_{p_1},[-1]_{p_2},[+1]_{p_3},[-1]_{p_4},[+1]_{q_1},[-1]_{q_2},[+1]_{q_3},[-1]_{q_4},\notag \\
&\hspace{2.5cm}[+1]_{r_1},[-1]_{r_2},[+1]_{r_3},[-1]_{r_4},[+1]_{s_1},[-1]_{s_2},[+1]_{s_3},[-1]_{s_4}),
\label{diaBC}
\end{align}
where $N=\sum_{k=1}^4(p_k+q_k+r_k+s_k)$, $0\leq p_k,q_k,r_k,s_k \leq N$ and $[+1]_{p_1}$s stand for
\begin{align}
[+1]_{p_1}=\underbrace{+1,\cdots ,+1}_{p_1}, \cdots ,
[-1]_{s_4}=\underbrace{-1,\cdots ,-1}_{s_4}.
\end{align}
Then the symmetry of BC becomes as
\begin{align}
&SU(N)~\longrightarrow ~SU(p_1)\times \cdots\times SU(p_4)\times SU(q_1)\times \cdots\times SU(q_4)\notag \\
&\hspace{0.5cm}\times SU(r_1)\times \cdots\times SU(r_4)\times SU(s_1)\times \cdots\times SU(s_4)\times U(1)^{15-l} ,
\label{br}
\end{align}
where $l$ is the number of $SU(0)$ and $SU(1)$ in $SU(p_1) \times \cdots \times SU(s_4)$.
Here and hereafter $SU(0)$ means nothing and $SU(1)$ unconventionally stands for $U(1)$.
We refer to BCs specified by diagonal matrices as {\it diagonal BC} 
and denote the above BC (\ref{diaBC}) as 
$[p_1,p_2,p_3,p_4;q_1,q_2,q_3,q_4;r_1,r_2,r_3,r_4;s_1,s_2,s_3,s_4]$. 
Note that the symmetry of BC is not necessarily identical to the physical symmetry.

Using the relations (\ref{R1}) and (\ref{R2}), we can derive the following equivalence relations:
\begin{align}
&[p_1,p_2,p_3,p_4;q_1,q_2,q_3,q_4;r_1,r_2,r_3,r_4;s_1,s_2,s_3,s_4] \notag\\
& \sim [p_1-1,p_2,p_3,p_4;q_1+1,q_2,q_3,q_4;r_1+1,r_2,r_3,r_4;s_1-1,s_2,s_3,s_4] \notag \\
&\hspace{10cm} ~\mathrm{for}~p_1,s_1\geq 1,\notag \\
&\hspace{0cm} \sim [p_1+1,p_2,p_3,p_4;q_1-1,q_2,q_3,q_4;r_1-1,r_2,r_3,r_4;s_1+1,s_2,s_3,s_4] \notag \\
&\hspace{10cm} ~\mathrm{for}~q_1,r_1\geq 1,\notag \\
&\hspace{0cm} \cdots \notag \\
&\hspace{0cm} \sim [p_1,p_2,p_3,p_4-1;q_1,q_2,q_3,q_4+1;r_1,r_2,r_3,r_4+1;s_1,s_2,s_3,s_4-1] \notag \\
&\hspace{10cm} ~\mathrm{for}~p_4,s_4\geq 1,\notag \\
&\hspace{0cm} \sim [p_1,p_2,p_3,p_4+1;q_1,q_2,q_3,q_4-1;r_1,r_2,r_3,r_4-1;s_1,s_2,s_3,s_4+1] \notag \\
&\hspace{10cm} ~\mathrm{for}~q_4,r_4\geq 1,\notag \\
& \sim [p_1-1,p_2+1,p_3+1,p_4-1;q_1,q_2,q_3,q_4;r_1,r_2,r_3,r_4;s_1,s_2,s_3,s_4] \notag \\
&\hspace{10cm} ~\mathrm{for}~p_1,p_4\geq 1,\notag \\
&\hspace{0cm} \sim [p_1+1,p_2-1,p_3-1,p_4+1;q_1,q_2,q_3,q_4;r_1,r_2,r_3,r_4;s_1,s_2,s_3,s_4] \notag \\
&\hspace{10cm} ~\mathrm{for}~p_2,p_3\geq 1,\notag \\
&\hspace{0cm} \cdots \notag \\
&\hspace{0cm} \sim [p_1,p_2,p_3,p_4;q_1,q_2,q_3,q_4;r_1,r_2,r_3,r_4;s_1-1,s_2+1,s_3+1,s_4-1] \notag \\
&\hspace{10cm} ~\mathrm{for}~s_1,s_4\geq 1,\notag \\
&\hspace{0cm} \sim [p_1,p_2,p_3,p_4;q_1,q_2,q_3,q_4;r_1,r_2,r_3,r_4;s_1+1,s_2-1,s_3-1,s_4+1] \notag \\
&\hspace{10cm} ~\mathrm{for}~s_2,s_3\geq 1. 
\end{align}
Hence the number of equivalence classes of BCs 
is ${}_{N+15}C_{15}-8\cdot {}_{N+13}C_{15}$.

When the BCs for bulk fields are given, mode expansions are carried out and the one-loop effective potential
for Wilson line phases is
calculated using the standard method.\cite{H,HHHK,HHK,KLY,HN&T,KK&M}
{}From the minimum of effective potential,
the physical symmetry and mass spectrum are obtained for each model.
We do not carry them out since our purpose is not to study the dynamics of models 
but to classify the BCs. 

As a comment, we can extend our argument to the case with the orbifold $S^1/Z_2\times \cdots \times S^1/Z_2$.
In this case, diagonal BCs are specified by $4^k$ kinds of integers and the number of equivalence classes 
is ${}_{N+4^k-1}C_{4^k-1}-4^{k-1}k\cdot{}_{N+4^k-3}C_{4^k-1}$, where $k$ is the number of $S^1/Z_2$.

\section{$T^2/Z_m$ Orbifold and Equivalence classes}

In this section, we study $SU(N)$ gauge theory on $M^4\times T^2/Z_m$ where $m = 2, 3, 4, 6$.
We discuss equivalence classes of BCs and obtain the number of BCs related to
diagonal representatives for each orbifold.

\subsection{$T^2/Z_2$ orbifold}

\begin{wrapfigure}{l}{6.6cm}
\label{F1.T2Z2}
\begin{center}
\unitlength 0.1in
\begin{picture}( 20.0000, 16.0000)(  6.3000,-20.6000)
%
\special{pn 13}%
\special{pa 1212 1774}%
\special{pa 2568 1768}%
\special{fp}%
\special{sh 1}%
\special{pa 2568 1768}%
\special{pa 2500 1748}%
\special{pa 2514 1768}%
\special{pa 2500 1788}%
\special{pa 2568 1768}%
\special{fp}%
%
\special{pn 13}%
\special{pa 1200 1760}%
\special{pa 1202 460}%
\special{fp}%
\special{sh 1}%
\special{pa 1202 460}%
\special{pa 1182 528}%
\special{pa 1202 514}%
\special{pa 1222 528}%
\special{pa 1202 460}%
\special{fp}%
%
\special{pn 13}%
\special{pa 2560 1760}%
\special{pa 2570 438}%
\special{dt 0.045}%
%
\special{pn 13}%
\special{pa 1190 440}%
\special{pa 2550 450}%
\special{dt 0.045}%
%
\special{pn 20}%
\special{sh 1}%
\special{ar 1212 1768 10 10 0  6.28318530717959E+0000}%
\special{sh 1}%
\special{ar 1212 1768 10 10 0  6.28318530717959E+0000}%
%
\special{pn 20}%
\special{sh 1}%
\special{ar 1200 1120 10 10 0  6.28318530717959E+0000}%
\special{sh 1}%
\special{ar 1200 1120 10 10 0  6.28318530717959E+0000}%
%
\special{pn 20}%
\special{sh 1}%
\special{ar 1890 1770 10 10 0  6.28318530717959E+0000}%
\special{sh 1}%
\special{ar 1890 1770 10 10 0  6.28318530717959E+0000}%
\put(9.6100,-19.4000){\makebox(0,0)[lb]{$O$}}%
\put(16.9000,-20.3000){\makebox(0,0)[lb]{$e_1/2$}}%
\put(7.500,-12.1000){\makebox(0,0)[lb]{$e_2/2$}}%
\put(26.3000,-18.3000){\makebox(0,0)[lb]{$e_1$}}%
\put(8.3000,-4.3000){\makebox(0,0)[lb]{$e_2$}}%
%
\special{pn 20}%
\special{sh 1}%
\special{ar 1880 1130 10 10 0  6.28318530717959E+0000}%
\special{sh 1}%
\special{ar 1880 1130 10 10 0  6.28318530717959E+0000}%
\put(15.4000,-10.7000){\makebox(0,0)[lb]{$(e_1+e_2)/2$}}%
\end{picture}%
\end{center}
\caption{Orbifold $T^2/Z_2$.}
\end{wrapfigure}
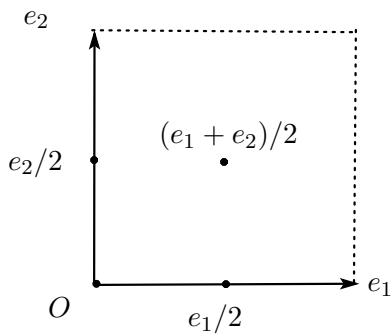
We study $SU(N)$ gauge theory on $M^4 \times T^2/Z_2$.\footnote{
Equivalence classes of BCs and dynamical gauge symmetry breaking were studied for $SU(2)$ gauge theory
on $T^2/Z_2$ in Ref.~\citen{HN&T}.}
Let $z$ be the complex coordinate of $T^2/Z_2$.
Here, $T^2$ is constructed by the $SU(2)\times SU(2)(\simeq SO(4))$ lattice.  
On $T^2$, the points $z+e_1$ and $z+e_2$ are identified with the point $z$ where 
$e_1$ and $e_2$ are basis vectors and we take $e_1 = 1$ and $e_2 = i$.
The orbifold $T^2/Z_2$ is obtained by further identifying $-z$ by $z$.
The resultant space is the area depicted in Fig.~1.
Fix points $z_{\tiny{\mbox{fp}}}$ for the $Z_2$ transformation $z \to \theta z=-z$ satisfy
\begin{align}
z_{\tiny{\mbox{fp}}} = \theta z_{\tiny{\mbox{fp}}} + n e_1 + n e_2 ,
\end{align}
where $m$ and $n$ are integers that characterized fixed points.
There are four kinds of points $0$, $e_1/2$, $e_2/2$, $(e_1+e_2)/ 2$.
Around these points, we define following six kinds of transformations:
\begin{align}
s_0&: z\rightarrow -z,~~
s_1: z\rightarrow -z+e_1, ~~
s_2: z\rightarrow -z+e_2, \notag \\
s_3&: z\rightarrow -z+e_1+e_2, ~~
t_1: z\rightarrow z+e_1,~~
t_2: z\rightarrow z+e_2.
\end{align}
These satisfy the following relations:
\begin{align}
&s_0^2=s_1^2=s_2^2=s_3^2=I,~s_1=t_1s_0,~s_2=t_2s_0,\notag \\
&s_3=t_1t_2s_0=s_1s_0s_2=s_2s_0s_1,~t_1t_2=t_2t_1.
\end{align}
The BCs of bulk fields are specified by matrices $(P_0,P_1,P_2,P_3,U_1,U_2)$
satisfying the relations: 
\begin{align}
&P_{0}^2=P_{1}^2=P_{2}^2=P_3^2=I,~P_{1}=U_1P_{0},~P_{2}=U_2P_{0},\notag \\
&P_3=U_1U_2P_0=P_{1}P_0P_{2}=P_{2}P_0P_{1},~U_1U_2=U_2U_1 .
\label{Z2-rel}
\end{align}
Because three of these matrices are independent,
we choose three kinds of unitary and hermitian matrices $P_0$, $P_1$ and $P_2$.

Given the BCs $(P_0,P_1,P_2)$, there still residual gauge invariance.
Under the gauge transformation with $\Omega(x,z,\bar{z})$, $P_0$, $P_1$ and $P_2$ are transformed as
\begin{align}
&P'_0(z, \bar{z})=\Omega(x,-z,-\bar{z})P_0\Omega^{\dag}(x,z,\bar{z}),\notag \\
&P'_1(z, \bar{z})=\Omega(x,-z+e_1,-\bar{z}+\bar{e}_1)P_1\Omega^{\dag}(x,z,\bar{z}),\notag \\
&P'_2(z, \bar{z})=\Omega(x,-z+e_2,-\bar{z}+\bar{e}_2)P_2\Omega^{\dag}(x,z,\bar{z}).
\end{align}
These BCs should be equivalent:
\begin{align}
(P_0,P_1,P_2) \sim (P'_0(z, \bar{z}),P'_1(z, \bar{z}),P'_2(z, \bar{z}))  .
\end{align}
This equivalence relation defines equivalence classes of the BCs.
Let us consider an $SU(2)$ gauge theory with 
the gauge transformation function defined by
\begin{align}
\Omega(z,\bar{z}) =\exp\left[i\alpha (a\tau_1+b\tau_2)z+i\bar{\alpha}(a\tau_1+b\tau_2)\bar{z}\right],
\label{OmegaZ2}
\end{align}
where $a$ and $b$ are real numbers.
When we take $(P_0,P_1,P_2)=(\tau_3,\tau_3,\tau_3)$, they are transformed as
\begin{align}
(\tau_3, \tau_3,\tau_3) \to 
(\tau_3, e^{2i{\tiny{\mbox{Re}}}\alpha(a\tau_1+b\tau_2)}\tau_3, e^{-2i{\tiny{\mbox{Im}}}\alpha(a\tau_1+b\tau_2)}\tau_3) .
\end{align}
In this way, we find the following equivalence relations among diagonal representatives:
\begin{align}
(\tau_3, \tau_3, \tau_3)\sim (\tau_3,\tau_3,-\tau_3) \sim (\tau_3, -\tau_3,\tau_3,) \sim (\tau_3, -\tau_3,-\tau_3).
\label{RZ2}
\end{align}

We carry out the classification of BCs for fields on the orbifold $T^2/Z_2$.
It is shown that all BCs are specified by diagonal matrices for $SU(2)$ gauge group.
As shown in Ref.~\citen{HN&T}, the $2 \times 2$ matrices that satisfy the relations (\ref{Z2-rel}) are given by
$(P_0,P_1,P_2) = (I, I, I)$, $(I, I, \tau_3)$, $(I,\tau_3,I)$,
$(I, \tau_3, \tau_3)$ and $(\tau_3, \tau_3 e^{2i{\tiny{\mbox{Re}}}\alpha(a\tau_1 + b\tau_2)}, 
\tau_3 e^{-2i{\tiny{\mbox{Im}}}\alpha(a\tau_1 + b\tau_2)})$ up to a sign factor for each component 
using global $SU(2)$ transformation.
In the case that $(P_0,P_1,P_2) = (\tau_3, \tau_3 e^{2i{\tiny{\mbox{Re}}}\alpha(a\tau_1 + b\tau_2)}, 
\tau_3 e^{-2i{\tiny{\mbox{Im}}}\alpha(a\tau_1 + b\tau_2)})$,
we obtain $(P_0,P_1,P_2) = (\tau_3, \tau_3, \tau_3)$ up to a sigh factor for each component 
after the gauge transformation with $\Omega(z,\bar{z})$ given by (\ref{OmegaZ2}).

For $SU(N)$ $(N \ne 2, 3)$ gauge group, there are BCs specified by matrices 
that cannot be diagonalized simultaneously by global unitary transformations and local gauge transformations. 
Here we give an example.
The following set of $4 \times 4$ matrices satisfy the relations (\ref{Z2-rel}),
\begin{align}
P_0 = 
\left(
\begin{array}{cc}
\tau_3 & 0  \\
0 & \tau_3 
\end{array}
\right) , ~~
P_1 = 
\left(
\begin{array}{cc}
\tau_2 & 0 \\
0 & -\tau_2 
\end{array}
\right) ,~~
P_2 = 
\frac{1}{\sqrt{2}}\left(
\begin{array}{cc}
\tau_2 e^{i\zeta \tau_1} & \tau_1 \\
\tau_1 & \tau_2 e^{i\zeta \tau_1}
\end{array}
\right) ,
\label{off-diag2}
\end{align}
where $\zeta$ is an arbitrary real number.
In the case with $\zeta \ne n\pi$ $(n \in \mathbb{Z})$, 
the above matrices (\ref{off-diag2}) cannot be transformed into diagonal ones simultaneously.
The symmetry of BC (\ref{off-diag2}) is nothing because there are no $4 \times 4$ traceless diagonal matrices 
simultaneously commutable to the above matrices.
Every component in $SU(N)$ multiplet does not necessarily become a simultaneous eigenstate of $Z_2$ parities 
if BCs contain off-diagonal elements.
As an example, we consider the $SU(N)$ gauge field $A_{M} = A_M^a T^a$ which satisfies 
the BC for the $Z_2$ transformation $z \to -z + e_2$,
\begin{align}
A_M(x, -z+e_2, -\bar{z}+\bar{e}_2)=\kappa_{[M]} P_{2}A_M(x, z, \bar{z})P_{2}^{\dag} ~,
\label{AM-P2} 
\end{align}
where $\kappa_{[\mu]}=1$, $\kappa_{[z]}=-1$ and $\kappa_{[\bar{z}]}=-1$.
In terms of components $A_M^a$, the above BC (\ref{AM-P2}) is written by
\begin{align}
& A_M^a(x, -z+e_2, -\bar{z}+\bar{e}_2)=\sum_b C^a_{[M]b} A_M^b(x, z, \bar{z}) ~,
\nonumber \\
& C^a_{[M]b} \equiv 2 \kappa_{[M]} \mbox{Tr}(P_{2}T^bP_{2}^{\dag}T^a) ~,
\label{AMa-P2} 
\end{align}
where we use the relation $\mbox{Tr}(T^a T^b)=\delta^{ab}/2$.
The $A_M^a$ do not have a definite $Z_2$ parity for the components 
which become mixed with others through the mixing matrices $C^a_{[M]b}$.
In this way, the reduction of rank can be done using BCs including off-diagonal elements and
this would be useful for the model-building of grand unification or gauge-Higgs unification.

Hereafter, we carry out the classification of BCs specified by diagonal matrices. 
The diagonal $P_0$, $P_1$ and $P_2$ are specified by eight non-negative integers $(p_i,q_i,r_i,s_i)$ $(i=1,2)$ such that
\begin{align}
&P_{0} = {\mbox{diag}}(\overbrace{[+1]_{p_1},[+1]_{p_2}}^p,\overbrace{[+1]_{q_1},[+1]_{q_2}}^q,\overbrace{[-1]_{r_1},[-1]_{r_2}}^r,\overbrace{[-1]_{s_1},[-1]_{s_2}}^{N-p-q-r}) , \nonumber \\
&P_{1} = {\mbox{diag}}([+1]_{p_1},[+1]_{p_2},[-1]_{q_1},[-1]_{q_2},[+1]_{r_1},[+1]_{r_2},[-1]_{s_1},[-1]_{s_2}) , \nonumber \\
&P_{2} = {\mbox{diag}}([+1]_{p_1},[-1]_{p_2},[+1]_{q_1},[-1]_{q_2},[+1]_{r_1},[-1]_{r_2},[+1]_{s_1},[-1]_{s_2}) ,
\end{align}
where $0 \leq p_i,q_i,r_i,s_i \leq N$.
Then the symmetry of BC becomes as
\begin{align}
SU(N)~\longrightarrow ~&SU(p_1)\times SU(p_2)\times SU(q_1)\times SU(q_2)\notag \\
&\hspace{1cm}\times SU(r_1)\times SU(r_2)\times SU(s_1)\times SU(s_2)\times U(1)^{7-l}.
\end{align}
We denote the above BC as $[p_1,p_2;q_1,q_2;r_1,r_2;s_1,s_2]$.
Using the relations (\ref{RZ2}), we can derive the following relations in $SU(N)$ gauge theory:
\begin{align}
&[p_1,p_2;q_1,q_2;r_1,r_2;s_1,s_2]\notag \\ 
&\sim [p_1-1,p_2+1;q_1,q_2;r_1,r_2;s_1+1,s_2-1],\notag \\ 
&\sim [p_1-1,p_2;q_1+1,q_2;r_1,r_2+1;s_1,s_2-1],\notag \\
&\sim [p_1-1,p_2;q_1,q_2+1;r_1+1,r_2;s_1,s_2-1],
\hspace{1.5cm}\mathrm{for}~p_1,s_2\geq 1,\\ 
&\sim [p_1+1,p_2-1;q_1,q_2;r_1,r_2;s_1-1,s_2+1],\notag \\ 
&\sim [p_1,p_2-1;q_1+1,q_2;r_1,r_2+1;s_1-1,s_2],\notag \\
&\sim [p_1,p_2-1;q_1,q_2+1;r_1+1,r_2;s_1-1,s_2],
\hspace{1.5cm}\mathrm{for}~p_2,s_1\geq 1,\\ 
&\sim [p_1+1,p_2;q_1-1,q_2;r_1,r_2-1;s_1,s_2+1],\notag \\ 
&\sim [p_1,p_2+1;q_1-1,q_2;r_1,r_2-1;s_1+1,s_2],\notag \\
&\sim [p_1,p_2;q_1-1,q_2+1;r_1+1,r_2-1;s_1,s_2],
\hspace{1.5cm}\mathrm{for}~q_1,r_2\geq 1,\\ 
&\sim [p_1+1,p_2;q_1,q_2-1;r_1-1,r_2;s_1,s_2+1],\notag \\ 
&\sim [p_1,p_2+1;q_1,q_2-1;r_1-1,r_2;s_1+1,s_2],\notag \\
&\sim [p_1,p_2;q_1+1,q_2-1;r_1-1,r_2+1;s_1,s_2],
\hspace{1.5cm}\mathrm{for}~q_2,r_1\geq 1.
\end{align}
Hence the number of equivalence classes of BCs related to diagonal representatives is ${}_{N+7}C_{7}-3\cdot{}_{N+5}C_{7}$.

\subsection{$T^2/Z_3$ Orbifold}

\begin{wrapfigure}{l}{6.6cm}
\label{F1.T2Z3}
\begin{center}
\unitlength 0.1in
\begin{picture}( 24.0000, 13.5000)( 13.0000,-18.8000)
%
\special{pn 13}%
\special{pa 2282 1704}%
\special{pa 3638 1698}%
\special{fp}%
\special{sh 1}%
\special{pa 3638 1698}%
\special{pa 3570 1678}%
\special{pa 3584 1698}%
\special{pa 3570 1718}%
\special{pa 3638 1698}%
\special{fp}%
%
\special{pn 13}%
\special{pa 2282 1698}%
\special{pa 1614 582}%
\special{fp}%
\special{sh 1}%
\special{pa 1614 582}%
\special{pa 1630 650}%
\special{pa 1640 628}%
\special{pa 1664 630}%
\special{pa 1614 582}%
\special{fp}%
%
\special{pn 13}%
\special{pa 3630 1698}%
\special{pa 2944 566}%
\special{dt 0.045}%
%
\special{pn 13}%
\special{pa 1612 574}%
\special{pa 2952 566}%
\special{dt 0.045}%
%
\special{pn 20}%
\special{sh 1}%
\special{ar 2282 1698 10 10 0  6.28318530717959E+0000}%
\special{sh 1}%
\special{ar 2282 1698 10 10 0  6.28318530717959E+0000}%
%
\special{pn 20}%
\special{sh 1}%
\special{ar 2268 948 10 10 0  6.28318530717959E+0000}%
\special{sh 1}%
\special{ar 2268 948 10 10 0  6.28318530717959E+0000}%
%
\special{pn 20}%
\special{sh 1}%
\special{ar 2930 1316 10 10 0  6.28318530717959E+0000}%
\special{sh 1}%
\special{ar 2930 1316 10 10 0  6.28318530717959E+0000}%
\put(21.0000,-18.5000){\makebox(0,0)[lb]{$O$}}%
\put(20.3000,-11.4000){\makebox(0,0)[lb]{$(e_1+2e_2)/3$}}%
\put(37.0000,-17.6000){\makebox(0,0)[lb]{$e_1$}}%
\put(14.0000,-6.000){\makebox(0,0)[lb]{$e_2$}}%
\put(26.0000,-15.6000){\makebox(0,0)[lb]{$(2e_1+e_2)/3$}}%
\end{picture}%
\end{center}
\caption{Orbifold $T^2/Z_3$.}
\end{wrapfigure}
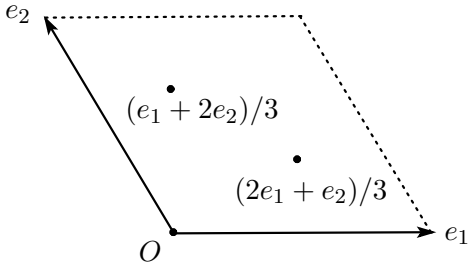
The BCs for $SU(N)$ gauge theory on $M^4 \times T^2/Z_3$
were studied in Ref.~\citen{KK&M}.\footnote{
The six-dimensional extension of $Z_3$ orbifold was initially introduced 
into the construction of four-dimensional heterotic string models.\cite{orbifold}. 
The models on $T^2/Z_3$ has been utilized in the search for the origin of three families\cite{threefamilies}
and the unification of gauge, Higgs and family.\cite{GLM&S}}
For completeness, we explain it briefly in this subsection.
Let $z$ be the coordinate of $T^2/Z_3$.
Here, $T^2$ is constructed by the $SU(3)$ lattice whose basis vectors are given 
by $e_1 = 1$ and $e_2 = e^{2\pi i/3} \equiv \omega$.
The orbifold $T^2/Z_3$ is obtained by further identifying $\omega z$ by $z$.
The resultant space is the area depicted in Fig.~2.
Fixed points for $Z_3$ transformation $z \to \omega z$
are $z=0$, $(2e_1 + e_2)/3$ and $(e_1 + 2 e_2)/3$.
Around these points, we define five kinds of transformations:
\begin{align}
& s_{0}: z \to \omega z,~~
s_{1}: z \to \omega z+e_{1} ,~~ s_{2}: z \to \omega z+e_{1}+e_{2} , \notag \\
& t_{1}: z \to z + e_{1}, ~~ t_{2}: z \to z + e_{2}.
\end{align}
Among the above operations, the following relations hold:
\begin{align}
& s_{0}^{3}=s_{1}^{3}=s_{2}^{3}
=s_{0}s_{1}s_{2} =s_{1}s_{2}s_{0} =s_{2}s_{0}s_{1}=I , \nonumber \\
& s_{1}=t_{1}s_{0},~~ s_{2}=t_{2}t_{1}s_{0}, ~~ t_{1}t_{2}=t_{2}t_{1}.
\end{align}
The BCs of bulk fields are specified by matrices $(\Theta_0,\Theta_1,\Theta_2,\Xi_1,\Xi_2)$
satisfying the relations: 
\begin{align}
&\Theta_0^3=\Theta_1^3=\Theta_2^3=\Theta_0\Theta_1\Theta_2
=\Theta_1\Theta_2\Theta_0=\Theta_2\Theta_0\Theta_1=I,\notag \\
&\Theta_1=\Xi_1\Theta_0,~~\Theta_2=\Xi_2\Xi_1\Theta_0,~~\Xi_1\Xi_2=\Xi_2\Xi_1 .
\label{RelZ3}
\end{align}
Because two of these matrices are independent,
we choose unitary matrices $\Theta_0$ and $\Theta_1$.

Given the BCs $(\Theta_0,\Theta_1)$, there still remains residual gauge invariance.
Under gauge transformation with $\Omega(x,z,\bar{z})$, $\Theta_0$ and $\Theta_1$ are transformed as
\begin{align}
& \Theta'_0(z, \bar{z}) = \Omega(x, \omega z, \bar{\omega} \bar{z}) \Theta_0 \Omega^{\dagger} (x, z, \bar{z}) , 
\nonumber \\
& \Theta'_1(z, \bar{z}) = \Omega(x, \omega z + 1, \bar{\omega} \bar{z} + 1) \Theta_1 \Omega^{\dagger} (x, z, \bar{z}) .
\end{align}
These BCs should be equivalent:
\begin{align}
(\Theta_0,\Theta_1) \sim (\Theta'_0(z, \bar{z}),\Theta'_1(z, \bar{z})) .
\end{align}
This equivalence relation defines equivalence classes of the BCs.

Let us consider an $SU(3)$ gauge theory with 
the gauge transformation function defined by
\begin{align}
\Omega(z,\bar{z}) = \exp\left[ia\left(Y^1_{+} z + Y^1_{-} \bar{z}\right)\right] ,
\end{align}
where $a$ is a real number and $Y^1_{+}$ and $Y^1_{-}$ are given by
\begin{align}
Y^1_{+} = 
\left(
\begin{array}{ccc}
0 & 1 & 0 \\
0 & 0 & 1 \\
1 & 0 & 0 
\end{array}
\right), ~~
Y^1_{-} = 
\left(
\begin{array}{ccc}
0 & 0 & 1 \\
1 & 0 & 0 \\
0 & 1 & 0 
\end{array}
\right),
\end{align}
When we take $(\Theta_0,\Theta_1)=(X,X)$ where $X$ is given by
\begin{align}
X = 
\left(
\begin{array}{ccc}
1 & 0 & 0 \\
0 & \omega & 0 \\
0 & 0 & \bar{\omega} 
\end{array}
\right) ,
\end{align}
they are transformed as
\begin{align}
(X, X) \to (X, e^{ia(Y^1_{+} + Y^1_{-})}X) .
\end{align}
In this way, we find the equivalence relations among diagonal representatives,
\begin{align}
(X,X) \sim (X,\bar{\omega}X) \sim (X,\omega X) ,
\label{RZ3}
\end{align}
where we use the relation
\begin{align}
\exp[iaY]= {1\over 3}(e^{2ia}+2e^{-ia})I+{1\over 3}(e^{2ia}-e^{-ia})Y .
\end{align}
Here, $I$ is the $3 \times 3$ unit matrix and $Y=Y^1_{+} + Y^1_{-}$.

There are BCs specified by matrices
that cannot be diagonalized simultaneously by global unitary transformations and local gauge transformations. 
For example, the following set of $3 \times 3$ matrices cannot be diagonalized simultaneously
by global unitary transformations and local gauge transformations,
\begin{align}
\Theta_0 = 
\left(
\begin{array}{ccc}
1 & 0 & 0 \\
0 & 1 & 0 \\
0 & 0 & 1
\end{array}
\right) , ~~
\Theta_1 = 
\left(
\begin{array}{ccc}
0 & e^{ia} & 0 \\
0 & 0 & e^{ib} \\
e^{ic} & 0 & 0
\end{array}
\right) ,~~
\Theta_2 = 
\left(
\begin{array}{ccc}
0 & 0 & e^{-ic} \\
e^{-ia} & 0 & 0 \\
0 & e^{-ib} & 0
\end{array}
\right) ~,
\label{off-diag3}
\end{align}
where $a$, $b$ and $c$ are arbitrary real numbers satisfying $a + b + c = 2n\pi$ ($n \in \mathbb{Z}$).
The above BC (\ref{off-diag3}) satisfies the relations (\ref{RelZ3}).
The symmetry of BC (\ref{off-diag3}) is nothing
because there are no $3 \times 3$ traceless diagonal matrices commutable to $(\Theta_1, \Theta_2)$ given in (\ref{off-diag3}).
The BCs specified by $N \times N$ matrices including off-diagonal elements can be constructed in the form that the above
set of $3 \times 3$ matrices or their transposed ones contain as submatrices.

We carry out the classification of BCs specified by diagonal matrices, for simplicity.
The diagonal $N \times N$ matrices $(\Theta_0, \Theta_1)$ are specified by nine non-negative integers $(p_j,q_j,r_j)$ $(j=1,2,3)$ such that
\begin{align}
&\Theta_0 = {\mbox{diag}}\overbrace{([1]_{p_1},[1]_{p_2},[1]_{p_3}}^p,
\overbrace{[\omega]_{q_1}, [\omega]_{q_2}, [\omega]_{q_3}}^q
\overbrace{[\bar{\omega}]_{r_1}, [\bar{\omega}]_{r_2}, [\bar{\omega}]_{r_3}}^{r=N-p-q}) , \nonumber \\
&\Theta_1 = {\mbox{diag}}([1]_{p_1}, [\omega]_{p_2}, [\bar{\omega}]_{p_3},[1]_{q_1}, [\omega]_{q_2}, 
[\bar{\omega}]_{q_3},[1]_{r_1}, [\omega]_{r_2}, [\bar{\omega}]_{r_3}) ,
\end{align}
where $0 \leq p,q,r \leq N$.
Then the symmetry of BC becomes as
\begin{align}
SU(N)\longrightarrow &SU(p_1)\times SU(p_2)\times SU(p_3)\times SU(q_1)\times SU(q_2)\notag \\
&~~~~~~~~~~~~~~~~~\times SU(q_3)\times SU(r_1)\times SU(r_2)\times SU(r_3)\times U(1)^{8-l}.
\end{align}
We denote the above BC as $[p_1,p_2,p_3;q_1,q_2,q_3;r_1,r_2,r_3]$.
Using the relations (\ref{RZ3}), we can derive the following equivalence relations in $SU(N)$ gauge theory:
\begin{align}
& [p_1, p_2, p_3; q_1, q_2, q_3; r_1, r_2, r_3]  
\nonumber \\ 
& \sim [p_1-1, p_2+1, p_3; q_1, q_2-1, q_3+1; r_1+1, r_2, r_3-1] , 
\nonumber \\
& \sim [p_1-1, p_2, p_3+1; q_1+1, q_2-1, q_3; r_1, r_2+1, r_3-1] , ~~
\mbox{for} ~~ p_1, q_2, r_3  \geq 1 , \\ 
&\sim [p_1+1, p_2-1, p_3; q_1, q_2+1, q_3-1; r_1-1, r_2, r_3+1] , 
\nonumber \\
&\sim [p_1, p_2-1, p_3+1; q_1+1, q_2, q_3-1; r_1-1, r_2+1, r_3] , ~~
\mbox{for} ~~ p_2, q_3, r_1  \geq 1 , \\
&\sim [p_1, p_2+1, p_3-1; q_1-1, q_2, q_3+1; r_1+1, r_2-1, r_3] , 
\nonumber \\
&\sim [p_1+1, p_2, p_3-1; q_1-1, q_2+1, q_3; r_1, r_2-1, r_3+1] , ~~
\mbox{for} ~~ p_3, q_1, r_2  \geq 1 .
\end{align}
Hence the number of equivalence classes of BCs related to diagonal representatives is ${}_{N+8}C_8-2\cdot{}_{N+5}C_8$.

\subsection{$T^2/Z_4$ Orbifold}

\begin{wrapfigure}{l}{6.6cm}
\label{F1.T2Z4}
\begin{center}
\unitlength 0.1in
\begin{picture}( 20.0000, 16.0000)(  6.3000,-20.6000)
%
\special{pn 13}%
\special{pa 1212 1774}%
\special{pa 2568 1768}%
\special{fp}%
\special{sh 1}%
\special{pa 2568 1768}%
\special{pa 2500 1748}%
\special{pa 2514 1768}%
\special{pa 2500 1788}%
\special{pa 2568 1768}%
\special{fp}%
%
\special{pn 13}%
\special{pa 1200 1760}%
\special{pa 1202 460}%
\special{fp}%
\special{sh 1}%
\special{pa 1202 460}%
\special{pa 1182 528}%
\special{pa 1202 514}%
\special{pa 1222 528}%
\special{pa 1202 460}%
\special{fp}%
%
\special{pn 13}%
\special{pa 2560 1760}%
\special{pa 2570 438}%
\special{dt 0.045}%
%
\special{pn 13}%
\special{pa 1190 440}%
\special{pa 2550 450}%
\special{dt 0.045}%
%
\special{pn 20}%
\special{sh 1}%
\special{ar 1212 1768 10 10 0  6.28318530717959E+0000}%
\special{sh 1}%
\special{ar 1212 1768 10 10 0  6.28318530717959E+0000}%
%
\special{pn 20}%
\special{sh 1}%
\special{ar 1200 1120 10 10 0  6.28318530717959E+0000}%
\special{sh 1}%
\special{ar 1200 1120 10 10 0  6.28318530717959E+0000}%
%
\special{pn 20}%
\special{sh 1}%
\special{ar 1890 1770 10 10 0  6.28318530717959E+0000}%
\special{sh 1}%
\special{ar 1890 1770 10 10 0  6.28318530717959E+0000}%
\put(9.6100,-19.4000){\makebox(0,0)[lb]{$O$}}%
\put(16.9000,-20.3000){\makebox(0,0)[lb]{$e_1/2$}}%
\put(8.3000,-12.1000){\makebox(0,0)[lb]{$e_2/2$}}%
\put(26.3000,-18.3000){\makebox(0,0)[lb]{$e_1$}}%
\put(9.300,-4.3000){\makebox(0,0)[lb]{$e_2$}}%
%
\special{pn 20}%
\special{sh 1}%
\special{ar 1880 1130 10 10 0  6.28318530717959E+0000}%
\special{sh 1}%
\special{ar 1880 1130 10 10 0  6.28318530717959E+0000}%
\put(15.4000,-10.7000){\makebox(0,0)[lb]{$(e_1+e_2)/2$}}%
\put(18.0700,-12.1000){\makebox(0,0)[lb]{~}}%
\put(11.3200,-18.4300){\makebox(0,0)[lb]{~}}%
\end{picture}%
\end{center}
\caption{Orbifold $T^2/Z_4$.}
\end{wrapfigure}
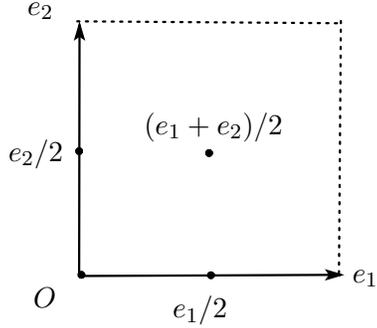
We study $SU(N)$ gauge theory on $M^4 \times T^2/Z_4$.
Let $z$ be the coordinate of $T^2/Z_4$.
Here, $T^2$ is constructed by $SU(2)\times SU(2)(\simeq SO(4))$ lattice
whose basis vectors are $e_1$ and $e_2$.
The orbifold $T^2/Z_4$ is obtained by further identifying $iz$ and $-z$ by $z$.
The resultant space is the area depicted in Fig.~3.
Then fixed points on $T^2/Z_4$ are $z=0$ and $(e_1+e_2)/2$ for $Z_4$ transformation $z \to iz$ 
and $z=0$, $e_1/2$, $e_2/2$ and $(e_1+e_2)/2$ for $Z_2$ transformation $z \to -z$.
Around these points, we define eight kinds of transformations:
\begin{align}
&s_0: z\rightarrow iz, ~~
s_1: z\rightarrow iz+e_1, ~~ s_{20}: z\rightarrow -z, \notag \\
&s_{21}: z\rightarrow -z+e_1, ~~
s_{22}: z\rightarrow -z+e_2, ~~
s_{23}: z\rightarrow -z+e_1+e_2, \notag \\
&t_1: z\rightarrow z+e_1, ~~
t_2: z\rightarrow z+e_2.
\end{align}
These satisfy the following relations:
\begin{align}
&s_0^4=s_1^4=s_{20}^2=s_{21}^2=s_{22}^2=s_{23}^2=I,~~s_1=t_1s_0,~~s_{21}=t_1s_{20}, \notag \\
&s_{22}=t_2s_{20},~~s_{20}=s_0^2,~~s_{21}=s_1s_0,~~s_{22}=s_0s_1, \notag \\
&s_{23}=t_1t_2s_{20}=s_{21}s_{20}s_{22}=s_{22}s_{20}s_{21},~~t_1t_2=t_2t_1.
\end{align}
The BCs of bulk fields are specified by matrices $(Q_0,Q_1,P_0,P_1,P_2,P_3,U_1,U_2)$
satisfying the relations: 
\begin{align}
&Q_0^4=Q_1^4=P_{0}^2=P_{1}^2=P_{2}^2=P_3^2=I,~Q_1=U_1Q_0,~P_{1}=U_1P_{0},\notag \\
&P_{2}=U_2P_{0},~P_0=Q_0^2,~P_1=Q_1Q_0,~P_2=Q_0Q_1,\notag \\
&P_3=U_1U_2P_0=P_{1}P_0P_{2}=P_{2}P_0P_{1},~U_1U_2=U_2U_1, 
\label{RelZ4}
\end{align}
where $Q_m$ $(m=0, 1)$ are unitary matrices and $P_n$ $(n=0,1,2,3)$ are unitary and hermitian matrices.
Because two of these matrices are independent,
we choose matrices $Q_0$ and $P_1$.

Given the BCs $(Q_0,P_1)$, there still remains residual gauge invariance. 
Under a gauge transformation $\Omega(x,z,\bar{z})$, $Q_0$ and $P_1$ are transformed as
\begin{align}
&Q'_0(z,\bar{z})=\Omega(x,iz,-i\bar{z})Q_0\Omega^{\dag}(x,z,\bar{z}), \notag \\
&P'_1(z,\bar{z})=\Omega(x,-z+e_1,-\bar{z}+\bar{e}_1)P_1\Omega^{\dag}(x,z,\bar{z}).
\end{align}
These BCs should be equivalent:
\begin{align}
(Q_0, P_1) \sim (Q'_0(z,\bar{z}), P'_1(z,\bar{z})) .
\end{align}
This equivalence relation defines equivalence classes of the BCs.
Let us consider an $SU(4)$ gauge theory with 
the gauge transformation function defined by
\begin{align}
\Omega(z,\bar{z})=\exp\{ia(Y_+^1 z+Y_-^1 \bar{z})\},
\end{align}
where $a$ is a real number and $Y_+^1$ and $Y_-^1$ are given by
\begin{align}
Y_+^1=\left(
\begin{array}{cccc}
0&-i&0&0\\0&0&i&0\\0&0&0&-i\\-i&0&0&0
\end{array}
\right) ,~~
Y_-^1=\left(
\begin{array}{cccc}
0&0&0&i\\i&0&0&0\\0&-i&0&0\\0&0&i&0
\end{array}
\right)  .
\end{align}
When we take $(Q_0,P_1)=(X,X^2)$ where $X$ is given by
\begin{align}
X=\left(
\begin{array}{cccc}
1& & &\mbox{\LARGE{0}}\\
&i& & \\
& &-1& \\
\mbox{\LARGE{0}}& & &-i
\end{array}
\right) ,
\end{align}
they are transformed as
\begin{align}
(X, X^2) \to (X, e^{ia(Y^1_{+} + Y^1_{-})}X^2) .
\end{align}
In this way, we find the following equivalence relation between diagonal representatives:
\begin{align}
(X, X^2) \sim (X, -X^2) ,
\label{RZ4}
\end{align}
where we use the relation
\begin{align}
\exp[iaY]= I \cos(\sqrt{2}a) + \frac{i}{\sqrt{2}}Y\sin(\sqrt{2}a)  .
\end{align}
Here $I$ is the $4 \times 4$ unit matrix and $Y=Y^1_{+} + Y^1_{-}$.

There are BCs specified by matrices 
that cannot be diagonalized simultaneously by global unitary transformations and local gauge transformations. 
For example, the following $4 \times 4$ matrices cannot be diagonalized simultaneously
by global unitary transformations and local gauge transformations,
\begin{align}
Q_0 = 
\left(
\begin{array}{cccc}
0 & e^{ia} & 0 & 0 \\
0 & 0 & e^{ib} & 0 \\
0 & 0 & 0 & e^{ic} \\
e^{id} & 0 & 0 & 0
\end{array}
\right) ~,~~
P_1 = 
\left(
\begin{array}{cccc}
1 & 0 & 0 & 0 \\
0 & 1 & 0 & 0 \\
0 & 0 & 1 & 0 \\
0 & 0 & 0 & 1
\end{array}
\right) ~,
\label{off-diag4}
\end{align}
where $a$, $b$, $c$ and $d$ are arbitrary real numbers satisfying $a + b + c +d= 2n\pi$ ($n \in \mathbb{Z}$).
The above BC (\ref{off-diag4}) satisfies the relations (\ref{RelZ4}).
The symmetry of BC (\ref{off-diag4}) is nothing
because there are no $4 \times 4$ traceless diagonal matrices commutable to $Q_0$ given in (\ref{off-diag4}).
The BCs specified by $N \times N$ matrices including off-diagonal elements can be constructed in the form that the above
set of $4 \times 4$ matrices or their transposed ones contain as submatrices.

We carry out the classification of BCs specified by diagonal matrices, for simplicity.
The diagonal $Q_0$ and $P_1$ are specified by eight non-negative integers $(p_i,q_i,r_i,s_i)$ $(i=1,2)$ such that
\begin{align}
Q_{0} &= {\mbox{diag}}(\overbrace{[+1]_{p_1},[+1]_{p_2}}^p,\overbrace{[+i]_{q_1},[+i]_{q_2}}^q,
\overbrace{[-1]_{r_1},[-1]_{r_2}}^r,\overbrace{[-i]_{s_1},[-i]_{s_2}}^{s=N-p-q-r}),\nonumber \\
P_{1} &= {\mbox{diag}}([+1]_{p_1},[-1]_{p_2},[+1]_{q_1},[-1]_{q_2},[+1]_{r_1},[-1]_{r_2},[+1]_{s_1},[-1]_{s_2}),
\end{align}
where $0 \leq p_i, q_i, r_i, s_i \leq N~(i=1,2)$. 
Then the symmetry of BC becomes as
\begin{align}
SU(N)~\longrightarrow ~&SU(p_1)\times SU(p_2)\times SU(q_1)\times SU(q_2) \notag \\
&\hspace{2cm}\times SU(r_1)\times SU(r_2)\times SU(s_1)\times SU(s_2)\times U^{7-l}.
\end{align}
We denote the above BC as $[p_1,p_2;q_1,q_2;r_1,r_2;$ $s_1,s_2]$.
Using the relation (\ref{RZ4}), we can derive the following equivalence relations in $SU(N)$ gauge theory:
\begin{align}
&[p_1,p_2;q_1,q_2;r_1,r_2;s_1,s_2;]\notag \\
&\hspace{1cm}\sim [p_1-1,p_2+1;q_1+1,q_2-1;r_1-1,r_2+1;s_1+1,s_2-1]\notag \\
&\hspace{9cm}\mathrm{for}~p_1,q_2,r_1,s_2 \geq 1 ,\notag \\
&\hspace{1cm}\sim [p_1+1,p_2-1;q_1-1,q_2+1;r_1+1,r_2-1;s_1-1,s_2+1]\notag \\
&\hspace{9cm}\mathrm{for}~p_2,q_1,r_2,s_1 \geq 1.
\end{align}
Hence the number of equivalence classes of BCs including diagonal representatives is ${}_{N+7}C_7-{}_{N+3}C_7$.

\subsection{$T^2/Z_6$ Orbifold}

\begin{wrapfigure}{l}{7.7cm}
\label{F1.T2Z6}
\begin{center}
\unitlength 0.1in
\begin{picture}( 30.1400, 14.5400)(2.900,-23.0000)
%
\special{pn 13}%
\special{pa 2506 2136}%
\special{pa 3308 2132}%
\special{fp}%
\special{sh 1}%
\special{pa 3308 2132}%
\special{pa 3240 2112}%
\special{pa 3254 2132}%
\special{pa 3240 2152}%
\special{pa 3308 2132}%
\special{fp}%
%
\special{pn 13}%
\special{pa 2506 2130}%
\special{pa 462 946}%
\special{fp}%
\special{sh 1}%
\special{pa 462 946}%
\special{pa 510 996}%
\special{pa 508 972}%
\special{pa 530 962}%
\special{pa 462 946}%
\special{fp}%
%
\special{pn 13}%
\special{pa 478 950}%
\special{pa 1282 956}%
\special{dt 0.045}%
%
\special{pn 20}%
\special{sh 1}%
\special{ar 2506 2132 10 10 0  6.28318530717959E+0000}%
\special{sh 1}%
\special{ar 2506 2132 10 10 0  6.28318530717959E+0000}%
%
\special{pn 20}%
\special{sh 1}%
\special{ar 1480 1540 10 10 0  6.28318530717959E+0000}%
\special{sh 1}%
\special{ar 1480 1540 10 10 0  6.28318530717959E+0000}%
%
\special{pn 20}%
\special{sh 1}%
\special{ar 2906 2134 10 10 0  6.28318530717959E+0000}%
\special{sh 1}%
\special{ar 2906 2134 10 10 0  6.28318530717959E+0000}%
\put(23.5000,-23.000){\makebox(0,0)[lb]{$O$}}%
\put(27.0000,-23.5000){\makebox(0,0)[lb]{$e_1/2$}}%
\put(12.0000,-18.0000){\makebox(0,0)[lb]{$e_2/2$}}%
\put(33.4400,-21.6700){\makebox(0,0)[lb]{$e_1$}}%
\put(3.3000,-9.1600){\makebox(0,0)[lb]{$e_2$}}%
%
\special{pn 20}%
\special{sh 1}%
\special{ar 2010 1550 10 10 0  6.28318530717959E+0000}%
\special{sh 1}%
\special{ar 2010 1550 10 10 0  6.28318530717959E+0000}%
\put(17.4900,-14.7200){\makebox(0,0)[lb]{$(e_1+e_2)/2$}}%
\put(10.5000,-14.13000){\makebox(0,0)[lb]{~}}%
\put(24.3000,-22.17000){\makebox(0,0)[lb]{~}}%
\put(17.6000,-18.26000){\makebox(0,0)[lb]{~}}%
\put(7.7000,-15.3000){\makebox(0,0)[lb]{$2e_2/3$}}%
\put(16.5000,-20.0000){\makebox(0,0)[lb]{$e_2/3$}}%
%
\special{pn 13}%
\special{pa 3304 2130}%
\special{pa 1276 956}%
\special{dt 0.045}%
\end{picture}%
\end{center}
\caption{Orbifold $T^2/Z_6$.}
\end{wrapfigure}
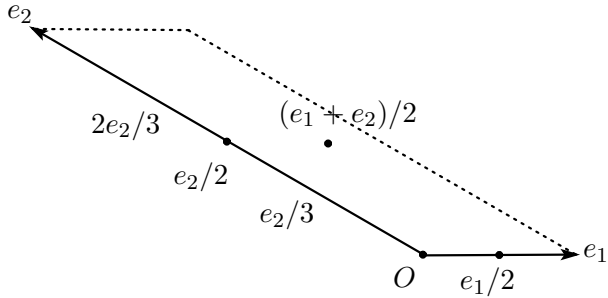
We study $SU(N)$ gauge theory on $M^4 \times T^2/Z_6$.
Let $z$ be the coordinate of $T^2/Z_6$.
Here, $T^2$ is constructed by the $G_2$ lattice
whose basis vectors are $e_1 = 1$ and $e_2 = (-3+i\sqrt{3})/2$.
The orbifold $T^2/Z_6$ is obtained by further identifying $\rho z$ by $z$ where $\rho^6 = 1$.
The resultant space is the area depicted in Fig.~4.
The $Z_6$ transformation $z \to \rho z$ is the $\pi/3$ rotation around the origin
and basis vectors are transformed as $\rho e_1=2e_1+e_2$, $\rho e_2=-3e_1-e_2$. 
Then fixed points on $T^2/Z_6$ are $z=0$ for $z \to \rho z$,
$z=0$, $e_2/3$ and $e_2/3$ for $z \to \rho^2 z$ 
and $z=0$, $e_1/2$, $e_2/2$ and $(e_1+e_2)/2$ for $z \to \rho^3 z$,
and around these points we define ten kinds of transformations:
\begin{align}
s_0&: z\rightarrow \rho z, ~~
s_{10}: z\rightarrow \rho^2 z, ~~
s_{11}: z\rightarrow \rho^2 z+e_1+e_2, ~~
s_{12}: z\rightarrow \rho^2 z+2e_1+2e_2, \notag \\
s_{20}&: z\rightarrow \rho^3 z, ~~
s_{21}: z\rightarrow \rho^3 z+e_1, ~~
s_{22}: z\rightarrow \rho^3 z+e_2, ~~
s_{23}: z\rightarrow \rho^3 z+e_1+e_2, \notag \\
t_1&: z\rightarrow z+e_1, ~~
t_2: z\rightarrow z+e_2.
\end{align}
Then satisfy the following relations;
\begin{align}
&s_0^6=s_{10}^3=s_{11}^3=s_{12}^3=s_{20}^2=s_{21}^2=s_{22}^2=s_{23}^2=I,
~s_{11}=t_1t_2s_{10},~s_{12}=t_1^2t_2^2s_{10},\notag \\
&s_{21}=t_1s_{20},~s_{22}=t_2s_{20},~s_{23}=t_1t_2s_{20}=s_{21}s_{20}s_{22}
=s_{22}s_{20}s_{21}=s_{11}s_0,\notag \\
&s_{10}=s_0^2,~s_{20}=s_0^3,~t_1t_2=t_2t_1.
\end{align}
The BCs of bulk fields are specified by matrices 
$(\Theta_{0},\Theta_{10},\Theta_{11},\Theta_{12}$, $\Theta_{20},\Theta_{21},\Theta_{22},\Theta_{23}$, $\Xi_1,\Xi_2)$
satisfying the relations: 
\begin{align}
&\Theta_{0}^6=\Theta_{10}^3=\Theta_{11}^3=\Theta_{12}^3=\Theta_{20}^2=\Theta_{21}^2=\Theta_{22}^2=\Theta_{23}^2=I,\notag \\
&\Theta_{11}=\Xi_1\Xi_2\Theta_{10},~\Theta_{12}=\Xi_1^2\Xi_2^2\Theta_{10},~\Theta_{21}=\Xi_1\Theta_{20},
~\Theta_{22}=\Xi_2\Theta_{20},\notag \\
&\Theta_{23}=\Xi_1\Xi_2\Theta_{20}=\Theta_{21}\Theta_{20}\Theta_{22}=\Theta_{22}\Theta_{20}\Theta_{21}
=\Theta_{11}\Theta_{0},\notag \\
&\Theta_{10}=\Theta_{0}^2,~\Theta_{20}=\Theta_{0}^3,~\Theta_{11}=\Theta_{23}\Theta_{20}\Theta_{10},
~\Theta_{12}=\Theta_{23}\Theta_{20}\Theta_{23}\Theta_{20}\Theta_{10},\notag \\
&\Xi_1\Xi_2=\Xi_2\Xi_1 .
\end{align}
Because three of these matrices are independent,
we choose unitary matrices $\Theta_{0}$, $\Theta_{21}$ and $\Theta_{22}$.

Given the BCs $(\Theta_{0},\Theta_{21},\Theta_{22})$, there still remains residual gauge invariance. 
Under a gauge transformation $\Omega(x,z,\bar{z})$, $\Theta_{0}$, $\Theta_{21}$ and $\Theta_{22}$ are transformed as
\begin{align}
& \Theta'_{0}(z,\bar{z}) = \Omega(x, \rho z, \bar{\rho} \bar{z}) \Theta_{0} \Omega^{\dagger} (x, z, \bar{z}) , 
\nonumber \\
& \Theta'_{21}(z,\bar{z})  = \Omega(x, \rho^3 z + e_1, \bar{\rho}^3 \bar{z} + \bar{e}_1) \Theta_{21} \Omega^{\dagger} (x, z, \bar{z}) ,
\nonumber \\
& \Theta'_{22}(z,\bar{z})  = \Omega(x, \rho^3 z + e_2, \bar{\rho}^3 \bar{z} + \bar{e}_2) \Theta_{22} \Omega^{\dagger} (x, z, \bar{z}) .
\end{align}
These BCs should be equivalent:
\begin{align}
(\Theta_{0},\Theta_{21},\Theta_{22}) \sim (\Theta'_{0}(z,\bar{z}) ,\Theta'_{21}(z,\bar{z}) ,\Theta'_{22}(z,\bar{z}) )  .
\end{align}
This equivalence relation defines equivalence classes of the BCs.
There are no equivalence relations between diagonal representatives.
To illustrate it,
let us consider an $SU(6)$ gauge theory with 
the gauge transformation function defined by
\begin{align}
\Omega(z,\bar{z})=\exp\{ia(Y_+^1 z+Y_-^1 \bar{z})\},
\end{align}
where $a$ is a real number and $Y_+^1$ and $Y_-^1=(Y_+^1)^{\dagger}$ are $6 \times 6$ matrices.
When  the diagonal $\Theta_{0}$ is transformed into the diagonal one $\Theta'_{0}$ under the gauge transformation, 
$\Theta_{0}$, $\Theta'_{0}$ and $Y_+^1$ are determined by
\begin{align}
\Theta_{0} = \Theta'_{0} = \left(
\begin{array}{cccccc}
1 & 0 & 0 & 0 & 0 & 0 \\
0 & \rho & 0 & 0 & 0 & 0 \\
0 & 0 & \rho^2 & 0 & 0 & 0 \\
0 & 0 & 0 & \rho^3 & 0 & 0 \\
0 & 0 & 0 & 0 & \rho^4 & 0 \\
0 & 0 & 0 & 0 & 0 & \rho^5
\end{array}
\right) , ~
Y_+^1=\left(
\begin{array}{cccccc}
0 & b_2 & 0 & 0 & 0 & 0\\
0 & 0 & b_3 & 0 & 0 & 0\\
0 & 0 & 0 & b_4 & 0 & 0\\
0 & 0 & 0 & 0 & b_5 & 0\\
0 & 0 & 0 & 0 & 0 & b_6\\
b_1 & 0 & 0 & 0 & 0 & 0
\end{array}
\right) ,
\end{align}
up to an overall factor $\rho^k$ for $\Theta_0$ and $\Theta'_0$.
Here $b_i$ $(i=1, \cdots, 6)$ are arbitrary complex numbers. 
It is shown that any diagonal $\Theta_{21}$ cannot be transformed into a different diagonal form.
Every diagonal representative is independent each other.
The diagonal $\Theta_{0}$, $\Theta_{21}$ and $\Theta_{22}$ for $SU(N)$ gauge theories are specified 
by twenty-four non-negative integers
and the number of equivalence classes of BCs related to diagonal representatives is ${}_{N+23}C_{23}$.

\section{Conclusions}

We have studied equivalence classes of BCs 
in an $SU(N)$ gauge theory on six-dimensional space-time including two-dimensional orbifold.
For five kinds of two-dimensional orbifolds $S^1/Z_2 \times S^1/Z_2$ and $T^2/Z_m$ $(m=2,3,4,6)$, 
orbifold conditions and those gauge transformation properties have been given 
and the equivalence relations among boundary conditions have been derived.
We have classified equivalence classes of BCs related to
diagonal representatives for each orbifold.     
There are BCs specified by matrices that cannot be diagonalized simultaneously
by global unitary transformations and local gauge transformations on $T^2/Z_m$.
Every component in $SU(N)$ multiplet does not necessarily become a simultaneous eigenstate for $Z_m$ transformations 
if BCs contain off-diagonal elements.
The reduction of rank can be done using these BCs and
this would be useful for the model-building of grand unification or gauge-Higgs unification.

If the BCs for bulk fields are given, mode expansions are carried out and the one-loop effective potential for Wilson line phases is
calculated using the standard method.
{}From the minimum of effective potential,
the physical symmetry and mass spectrum are obtained for each model.
It is crucial to study dynamical gauge symmetry breaking and mass generation
in a realistic model including fermions.
It is also important to construct a phenomenologically viable model realizing gauge-Higgs unification\cite{GHU} 
and/or family unification\cite{OFU} based on them.
The local grand unification can be realized by taking nontrivial BCs.\footnote{
The $\lq$local' gauge groups at fixed points were realized on $T^2/Z_2$ in Ref.~\citen{ABC}.
The string-derived orbifold grand unification theories were studied in Refs.~\citen{KRZ} and \citen{BHLR}.}
It is interesting to study the phenomenological aspects of such models.
We hope to further study these subjects in the near future.

\section*{Acknowledgements}
This work was supported in part by Scientific Grants from the Ministry of Education, Culture, Sports, Science and Technology
under Grant Nos.~18204024 and 18540259 (Y.~K.).


\end{document}